\documentclass[aps,twocolumn,showpacs]{revtex4}
\usepackage{graphicx}
\usepackage{epsfig}
\begin{document}

\title{Orbital mechanism of the circular photogalvanic effect in quantum wells}
\author{S.\,A.\,Tarasenko}
\affiliation{A.F.~Ioffe Physico-Technical Institute, Russian
Academy of Sciences, 194021 St.~Petersburg, Russia}

\begin{abstract}
It is shown that the free-carrier (Drude) absorption of circularly
polarized radiation in quantum well structures leads to an
electric current flow. The photocurrent reverses its direction
upon switching the light helicity. A pure orbital mechanism of
such a circular photogalvanic effect is proposed that is based on
interference of different pathways contributing to the light
absorption. Calculation shows that the magnitude of the helicity
dependent photocurrent in $n$-doped quantum well structures
corresponds to recent experimental observations.
\end{abstract}




\maketitle

\section{Introduction}

The absorption of circularly polarized light in semiconductor
structures may lead to generation of an electric current which
reverses its direction upon switching the light polarization from
right-handed to left-handed and vice versa. Such a circular
photogalvanic effect (CPGE) caused by asymmetry of elementary
processes of photoexcitation requires structures of the
appropriate symmetry. It is possible only in media which belong to
one of the gyrotropic classes~\cite{Sturman92,Ivchenko05}.

Recently, CPGE has attracted a great deal of attention. It was
observed in low-dimensional structures: in zinc-blende- and
diamond-type quantum wells (QWs), which are also belong to the
gyrotropic media~\cite{GanichevPrettl03}. Circular photocurrents
were studied for a variety of optical ranges, including interband
transitions in QWs~\cite{Belkov03,Bieler05,Yang06}, direct
intersubband~\cite{Ganichev03} and indirect intrasubband
(Drude-like)~\cite{Ganichev01} transitions in $n$-doped QW
structures.

So far, the microscopic theory of CPGE in zinc-blende-type
structures has been developed for the direct interband or
intersubband optical transitions. It has been shown that the
effect originates from the complicated band structure of the
semiconductor compounds, mostly, from spin-orbit splitting of the
electron or hole states, which is odd in the wave vector
$\mathbf{k}$~\cite{Geller89,Golub02,Ganichev03}, and
linear-in-$\mathbf{k}$ terms in the matrix elements of interband
optical transitions~\cite{Geller89,Khurgin06}. Here we address
CPGE caused by the free-carrier absorption of circularly polarized
radiation in $n$-doped QW structures. We show that a pure orbital
mechanism of the effect can predominate in this spectral range,
which is related to neither spin-orbit coupling nor spin-sensitive
selection rules for the optical transitions. This mechanism is
based on interference of different pathways contributing to the
light absorption. A similar CPGE model for the optical transitions
between quantum subbands was proposed in Ref.~\cite{Magarill82}.
However, the orbital mechanism is especially efficient for the
free-carrier absorption, which is contributed by indirect optical
transitions.

We consider CPGE in (001)-oriented QWs grown from zinc-blende-type
semiconductors. Symmetry analysis shows that the helicity
dependent photocurrent $\mathbf{j}$ in such structures can be
induced only at oblique incidence of the
radiation~\cite{Ivchenko05,GanichevPrettl03} and is described by
two linearly independent constants $\gamma_1$ and $\gamma_2$ as
follows:
\begin{eqnarray}\label{j_phenom}
j_x &=& [\gamma_2 \, l_x - \gamma_1 \, l_y] \,I P_{cicr} \:,\\
j_y &=& [\gamma_1 \, l_x - \gamma_2 \, l_y] \,I P_{cicr} \:,
\nonumber
\end{eqnarray}
where $\mathbf{l}=\mathbf{q}/q$ is the unit vector pointing in the
light propagation direction, $\mathbf{q}$ and $I$ are the wave
vector and intensity of the light inside the structure,
respectively; $P_{circ}$ is the radiation helicity ranging from
$-1$ (for the left-handed circular polarization) to $+1$ (for the
right-handed circular polarization); $x\parallel[100]$,
$y\parallel[010]$, and $z\parallel[001]$ are the cubic
crystallographic axes. Phenomenologically, the constant $\gamma_1$
is related to the heterostructure asymmetry, i.e., nonequivalence
of the $z$ and $-z$ directions, while the constant $\gamma_2$
originates from lack of an inversion center in the host crystal.

At normal incidence of the light the helicity dependent
photocurrent can be induced only in quantum wells grown along one
of the low-symmetric crystallographic directions, such as (110)-
or (113)-oriented structures. In this geometry CPGE is described
by
\begin{equation}\label{j_110}
j_{x'} = \gamma_3 \, l_{z'} I P_{cicr} \:,
\end{equation}
where $x'$ is an in-plane axis, which is pointed along
$[1\bar{1}0]$ for (110)- and (113)-oriented QWs, and $z'$ is the
growth direction. The phenomenological constant $\gamma_3$ is
entirely related to absence of an inversion center in the host
crystal, similarly to $\gamma_2$ in Eq.~(\ref{j_phenom}).

Below we present a microscopic theory for the orbital mechanism of
CPGE and show that this mechanism contributes to $\gamma_1$ as
well as $\gamma_2$ and $\gamma_3$ constants.

\section{Microscopic theory}

The free-carrier absorption of radiation is always accompanied by
electron scattering from static defects, acoustic or optical
phonons, etc., because of the need for energy and momentum
conservation. Such indirect optical transitions are treated as
second-order processes, which involve electron-photon interaction
and electron scattering, via virtual intermediate states. The
intermediate states can be those within the same quantum subband,
$e1$ in our case, or in other conduction or valence subbands. The
dominant pathway determining the QW absorbance involve
intermediate states within the subband $e1$ (see Fig.~1). The
matrix element of such kind of processes has the
form~\cite{Ivchenko05}
\begin{equation}\label{Me1}
M_{\mathbf{k}'\mathbf{k}}^{(e1)} = \frac{eA}{c\omega m^*}\,
\mathbf{e}\cdot(\mathbf{k}'-\mathbf{k}) \,V_{11} \:.
\end{equation}
Here $\mathbf{k}$ and $\mathbf{k}'$ are the initial and final
electron wave vectors, $e$ is the electron charge, $A$ is the
amplitude of the electromagnetic wave related to the light
intensity by $I=A^2 \omega^2 n_{\omega}/(2\pi c)$, $\omega$ is the
light frequency, $m^*$ is the effective electron mass,
$n_{\omega}$ is the refractive index of the medium, $\mathbf{e}$
is the (complex) unit vector of the light polarization, and
$V_{11}$ is the matrix element of intrasubband scattering. As
follows from Eq.~(\ref{Me1}), optical transitions with
intermediate states in the subband $e1$ can be induced only by
radiation with nonzero in-plane component of the polarization
vector $\mathbf{e}$. They do not lead to any circular photocurrent
because the square of the matrix element~(\ref{Me1}), which
determines the transition rate, is independent of the light
helicity.

\begin{figure}
  \includegraphics[height=.18\textheight]{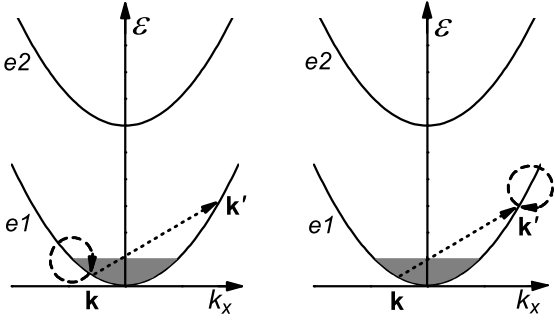}
  \caption{Intrasubband optical transitions
  $(e1,\mathbf{k}) \rightarrow (e1,\mathbf{k}')$
with intermediate states in the subband $e1$. Dashed circles and
dotted lines represent electron-photon interaction and electron
scattering, respectively. Panels correspond to two possible
processes: electron-photon interaction followed by electron
scattering (left panel) and electron scattering followed by
electron-photon interaction (right panel).}
\end{figure}
\begin{figure}
  \includegraphics[height=.18\textheight]{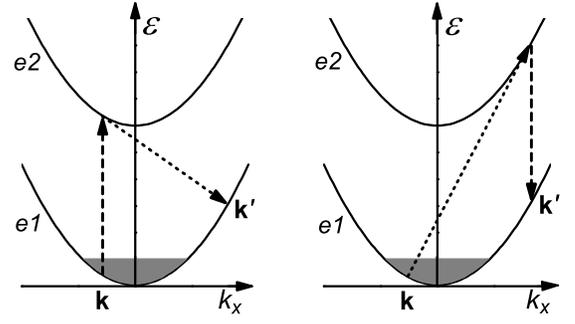}
  \caption{Intrasubband optical transitions
$(e1,\mathbf{k}) \rightarrow (e1,\mathbf{k}')$ with intermediate
states in the subband $e2$. Dashed and dotted lines represent
electron-photon interaction and electron scattering, respectively.
Panels correspond to two possible processes: electron-photon
interaction followed by electron scattering (left panel) and
electron scattering followed by electron-photon interaction (right
panel).}
\end{figure}

The helicity dependent photocurrent arises if one takes into
account interference of the processes depicted in Fig.~1 and those
with virtual intermediate states in other subbands.

\subsection*{Contribution to $\gamma_1$}

The quantum subband, which is the closest to $e1$, is the electron
subband $e2$. Optical transitions with intermediate states in the
subband $e2$ contributing to the free-carrier absorption are
sketched in Fig.~2. The matrix element of these indirect processes
has the form
\begin{equation}\label{Me2}
M_{\mathbf{k}'\mathbf{k}}^{(e2)} = i \frac{eA}{c \hbar} \left(
\frac{\varepsilon_{21}}{\varepsilon_{21}-\hbar\omega} -
\frac{\varepsilon_{21}}{\varepsilon_{21}+\hbar\omega} \right)
z_{21} \,e_z V_{21} \:.
\end{equation}
Here $\varepsilon_{21}$ is the energy separation between the
subbands $e2$ and $e1$, $z_{21}$ is the coordinate matrix element
between the envelope functions in the subbands, $\varphi_{1}(z)$
and $\varphi_{2}(z)$,
\begin{equation}\label{z21}
z_{21} = \int\limits_{-\infty}^{+\infty} \varphi_{2}(z) z \,
\varphi_{1}(z)\: dz \:,
\end{equation}
and $V_{21}$ is the matrix element of \textit{inter}-subband
scattering. In contrast to Eq.~(\ref{Me1}), the matrix
element~(\ref{Me2}) is nonzero only for radiation with
nonvanishing out-of-plane component of the polarization vector
$\mathbf{e}$ and contains the imaginary unit as a prefactor.

Light absorption by free carriers is contributed by processes with
all possible intermediate states. Making allowance for the
transitions via the subbands $e1$ and $e2$ one can write for the
photocurrent
\begin{equation}\label{j_gen}
\mathbf{j} = e \frac{4\pi}{\hbar} \sum_{\mathbf{k},\mathbf{k}'}
[\tau_p(\varepsilon_{\mathbf{k}'})\, \mathbf{v}_{\mathbf{k}'} -
\tau_p(\varepsilon_{\mathbf{k}})\, \mathbf{v}_{\mathbf{k}}]
\end{equation}
\vspace{-0.4cm}
\[
\times \left| M_{\mathbf{k}'\mathbf{k}}^{(e1)} +
M_{\mathbf{k}'\mathbf{k}}^{(e2)} \right|^2
(f_{\mathbf{k}}-f_{\mathbf{k}'})
\:\delta(\varepsilon_{\mathbf{k}'}-\varepsilon_{\mathbf{k}}-\hbar\omega)
\:,
\]
where $\mathbf{v}_{\mathbf{k}}=\hbar \mathbf{k}/m^*$ and
$\varepsilon_{\mathbf{k}}=\hbar^2 k^2/ (2m^*)$ are the electron
velocity and kinetic energy, respectively,
$\tau_p(\varepsilon_{\mathbf{k}})$ is the momentum relaxation
time, which may depend on the electron energy, $f_{\mathbf{k}}$ is
the function of equilibrium carrier distribution in the subband
$e1$, and the factor 4 in Eq.~(\ref{j_gen}) accounts for the spin
degeneracy.

The square of the matrix element sum in Eq.~(\ref{j_gen}) contains
the interference term
$2\mathrm{Re}[M_{\mathbf{k}'\mathbf{k}}^{(e1)}
M_{\mathbf{k}'\mathbf{k}}^{(e2)*}]$, which is odd in the wave
vector. Therefore, it leads to an asymmetrical distribution of the
photoexcited carriers in $\mathbf{k}$-space, i.e., to an electric
current. Moreover, the interference term is proportional to
components of $i[\mathbf{e}\times\mathbf{e}^*]$. It vanishes for
the linearly polarized light and is proportional to the light
helicity $P_{circ}$ for the elliptical or circular polarization,
because $i[\mathbf{e}\times\mathbf{e}^*]=\mathbf{l}P_{circ}$.
Thus, the sign of the interference term is determined by the light
helicity, and the photocurrent reverses its direction by changing
the light polarization from right-handed to left-handed and vice
versa.

Physically, the orbital mechanism of the circular photogalvanic
effect can also be interpreted as follows. Under excitation with
circularly polarized light the processes depicted in Fig.~1 and
Fig.~2 are added constructively for transitions to positive $k_x'$
(or negative $k_x'$, depending on the light helicity) and
destructively for transitions to $-k_x'$. It means that the
transition rates to the states $k_x'$ and $-k_x'$ are different,
resulting in an electric current of the photoexcited carriers.
Direction of the photocurrent with respect to the crystallographic
axes is determined by the light polarization and the explicit form
of the matrix elements of intrasubband and intersubband
scattering.

The matrix element~(\ref{Me2}) is proportional to out-of-plane
component of the polarization vector. Therefore, the term caused
by interference of the optical processes via the subband $e1$ and
$e2$ can contribute to CPGE at oblique incidence only. We consider
the effect for $(001)$-grown quantum wells. In such structures
electron scattering by static defects, acoustic or optical phonons
is usually treated as central. Under this assumption the matrix
elements of intrasubband and intersubband scattering, $V_{11}$ and
$V_{21}$, respectively, take no account of the lattice symmetry of
the host semiconductor and depend on $|\mathbf{k}'-\mathbf{k}|$
only. Then, the helicity dependent photocurrent flows in the
direction perpendicular to the light incidence plane. This
contribution corresponds to the terms described by the
phenomenological constant $\gamma_1$ in Eq.~(\ref{j_phenom}).
Calculation after Eq.~(\ref{j_gen}) shows that, in the case of
elastic scattering and provided $\hbar\omega \ll
\varepsilon_{21}$, the constant is given by
\begin{equation}\label{gamma1}
\gamma_1 = - 2 e \tau_p \frac{\omega z_{21}}{\varepsilon_{21}}
\,\xi \eta_{\|} \:,
\end{equation}
where $\xi$ is a dimensionless parameter, which depends on the
structure design and mechanisms of scattering, and $\eta_{\|}$ is
the QW absorbance for radiation polarized in the QW plane. We note
that both the absorbance $\eta_{\|}$ and the momentum relaxation
time $\tau_p$ are governed by the matrix element of intrasubband
scattering $V_{11}$. Thus, the product $\tau_p \eta_{\|}$ is
independent of the scattering rate. For quasi-elastic scattering
from acoustic phonons or short-range static defects, the product
is given by
\begin{equation}\label{tau*eta}
\tau_p \,\eta_{\|} = \frac{2\pi \alpha}{n_{\omega}} \frac{\kappa
\hbar}{m^* \omega^2} N_e  \:,
\end{equation}
where $\alpha=e^2/\hbar c \approx1/137$ is the fine-structure
constant, $N_e$ is the carrier density, and $\kappa$ is a
parameter depending on the carrier distribution. The latter is
equal to $1$ and $2$ for the cases $\hbar\omega \gg
\bar{\varepsilon}$ and $\hbar\omega \ll \bar{\varepsilon}$,
respectively, with $\bar{\varepsilon}$ being the mean kinetic
energy of equilibrium carriers ($\bar{\varepsilon}=E_F/2$ for a
degenerate two-dimensional gas and $\bar{\varepsilon}=k_BT$ for a
non-degenerate two-dimensional gas, where $E_F$ is the Fermi
energy, $k_B$ is the Boltzmann constant, and $T$ is the
temperature).

At low temperatures the free-carrier absorption is dominated by
processes, which involve elastic scattering from static defects
such as impurities, imperfections of the QW interfaces, etc. For
electron scattering from short-range defects the parameter $\xi$
has the form
\begin{equation}\label{xi}
\xi = \left. \int\limits_{-\infty}^{+\infty} \varphi_{1}^3(z)
\varphi_{2}(z) w(z) dz \right/ \int\limits_{-\infty}^{+\infty}
\varphi_{1}^4(z) w(z) dz  \:,
\end{equation}
where $w(z)$ is the  distribution function of the scatterers along
the growth direction. At higher temperatures electron-phonon
interaction predominates over the electron scattering by static
defects. In the case of quasi-elastic scattering from acoustic
phonons the parameter $\xi$ is also given by Eq.~(\ref{xi}) where
$w(z)$ is set to be a constant.

In accordance with general symmetry arguments, the helicity
dependent photocurrent corresponding to the phenomenological
constant $\gamma_1$ is related to inversion asymmetry of the
heterostructure. This follows also from Eqs.~(\ref{gamma1}) and
(\ref{xi}), which demonstrate that the sign and magnitude of
$\gamma_1$ is determined by asymmetry of the confinement potential
and the doping profile. In particular, $\gamma_1\equiv0$ for the
absolutely symmetrical structures, where $w(z)$ and $\varphi_1(z)$
are even functions while $\varphi_2(z)$ is an odd function with
respect to the QW center.

\subsection*{Contribution to $\gamma_2$ and $\gamma_3$}

To obtain the circular photocurrents described by the
phenomenological constants $\gamma_2$ in Eq.~(\ref{j_phenom}) and
$\gamma_3$ in Eq.~(\ref{j_110}) one has to take into account the
lattice symmetry of the QW host semiconductor. This can be done
considering interference of optical transitions with intermediate
states in the subband $e1$ and those via the valence-band states.

To calculate this contribution to CPGE we neglect spin-orbit
splitting of the valence band for simplicity and assume that the
effective hole masses in the QW plane are larger than the
effective electron mass. Then, the matrix element of the
intrasubband optical transitions via the valence-band states can
be presented as
\begin{equation}\label{Mh1}
M_{\mathbf{k}'\mathbf{k}}^{(v)} = i \frac{eA}{c \hbar}
\frac{\hbar\omega}{E_g^2} \,P \sum_{j} e_{j} V_{S R_{j}} \:.
\end{equation}
Here $E_g$ is the band gap energy, $P=i(\hbar/m_0) \langle S | p_z
| R_z \rangle$ is the Kane matrix element, $m_0$ is the free
electron mass, $S$ and $R_j$ ($j=x,y,z$) are the Bloch functions
of the conduction and valence bands at $\Gamma$-point of the
Brillouin zone, respectively, and $V_{S R_{j}}$ are the matrix
elements of \textit{inter}-band scattering. For scattering from
acoustic phonons the matrix elements $V_{S R_{j}}$ have the
form~\cite{optor,Ivchenko04}
\begin{eqnarray}
V_{S R_x} = \Xi_{cv} \langle \mathbf{k}' | u_{yz} | \mathbf{k}
\rangle
\:, \\
V_{S R_y} = \Xi_{cv} \langle \mathbf{k}' | u_{xz} | \mathbf{k}
\rangle
\nonumber \:, \\
V_{S R_z} = \Xi_{cv} \langle \mathbf{k}' | u_{xy} | \mathbf{k}
\rangle \nonumber \:,
\end{eqnarray}
where $\Xi_{cv}$ is the interband deformation-potential constant,
$\langle \mathbf{k}' | u_{\alpha\beta} | \mathbf{k} \rangle$ are
the matrix elements of coupling between the electron states
$\mathbf{k}$ and $\mathbf{k}'$ in the subband $e1$ caused by the
deformation, and $u_{\alpha\beta}$ ($\alpha \neq \beta$) are the
phonon-induced off-diagonal components of the strain tensor. The
interband constant $\Xi_{cv}$ originates from lack of an inversion
symmetry in zinc-blende-type crystals ($T_d$ point group) and
vanishes in centrosymmetric semiconductors~\cite{optor}.

Calculation shows that in (001)-oriented QWs the matrix element
$M_{\mathbf{k}'\mathbf{k}}^{(v)}$ contains a term proportional to
$(k_x'-k_x)(k_y'-k_y)e_z$. Interference of this term and
$M_{\mathbf{k}'\mathbf{k}}^{(e1)}$ gives rise to the circular
photocurrent described by $\gamma_2$ in Eq.~(\ref{j_phenom}). For
the free-carrier absorption assisted by quasi-elastic scattering
from acoustic phonons the constant $\gamma_2$ has the form
\begin{equation}\label{gamma2}
\gamma_2 = - 2 e \tau_p \frac{\Xi_{cv}}{\Xi_c} \frac{\omega
P}{E_g^2} \,\zeta \eta_{\|} \:,
\end{equation}
where $\Xi_c$ is the intraband deformation-potential constant,
which determines the light absorbance $\eta_{\|}$, and $\zeta$ is
a parameter depending on the QW width, the carrier distribution
and the photon energy. For a rectangular quantum well with the
infinitely high barriers, $\zeta=8\pi \bar{k}a/3$ if the photon
energy is much less than the mean electron energy and
$\zeta=k_{\omega}a/12$ in the opposite limiting case. Here
$\bar{k}$ is the mean value of $|\mathbf{k}|$ ($\bar{k}=2k_F/3$
for a degenerate two-dimensional gas and $\bar{k}=\sqrt{\pi m^*
k_B T/(2\hbar^2)}$ for a non-degenerate two-dimensional gas, where
$k_F$ is the Fermi wave vector),
$k_{\omega}=\sqrt{2m^*\omega/\hbar}$, and $a$ is the QW width.

In (110)-oriented QWs interference of the optical transitions with
intermediate states in the subband $e1$ and in the valence band
leads to the helicity dependent photocurrent even at normal
incidence of the light. Calculation shows that for this particular
mechanism the constant $\gamma_3$ has the form
\begin{equation}\label{gamma3}
\gamma_3 = - \frac{\gamma_2}{4\zeta} \:.
\end{equation}
It is independent of the QW width and is determined solely by the
carrier density, the scattering mechanism and the band parameters.

As is mentioned above, the phenomenological constants $\gamma_2$
and $\gamma_3$ are related to lack of an inversion asymmetry in
the bulk semiconductor rather than to the structure asymmetry.
This follows also from Eqs.~(\ref{gamma2}) and (\ref{gamma3}),
which show that both $\gamma_2$ and $\gamma_3$ do not vanish even
in symmetrical QWs grown from zinc-blende-type compounds.

Another orbital contribution to $\gamma_2$ can originate from
interference of the processes via the subbands $e1$ and $e2$
(depicted in Figs.~1 and 2, respectively) in the presence of
scattering anisotropy. Indeed, the $T_d$ point group of
zinc-blende-type semiconductors does not contain the space
inversion and allows for an antisymmetric term in the matrix
element of scattering, which changes the sign upon the replacement
$\mathbf{k} \rightarrow -\mathbf{k}$ and $\mathbf{k}' \rightarrow
-\mathbf{k}'$. Such an anisotropy in scattering can be modeled,
e.g., by impurity potential of the form~\cite{Levitov85}
\begin{equation}\label{potential}
U_i(\mathbf{r})=
U_0(|\mathbf{r}-\mathbf{r}_i|)+U_3(|\mathbf{r}-\mathbf{r}_i|)(x-x_i)(y-y_i)(z-z_i)
\:,
\end{equation}
where $\mathbf{r}_i$ is the impurity position. The antisymmetric
part of the impurity potential does not violate inherent isotropy
of electrical properties of bulk cubic materials and, therefore,
can not be easily detected~\cite{scattering}. However, in
(001)-oriented QWs it adds a term to the matrix element of
electron scattering, which is proportional to
$(k_x'-k_x)(k_y'-k_y)$ and, therefore, contributes to $\gamma_2$.
We assume that the impurity range is much shorter than the QW
width and the inequality $\hbar\omega \ll \varepsilon_{21}$ is
fulfilled. Then, the contribution to $\gamma_2$ caused by the
scattering anisotropy has the form
\begin{equation}\label{gamma2b}
\gamma_2' = e \tau_p  \frac{u_3}{u_0} \frac{\omega
z_{21}}{\varepsilon_{21}} \frac{m^* E}{\hbar^2} \,Q \eta \:,
\end{equation}
where $u_0$ and $u_3$ are related to the symmetric and
antisymmetric parts of the impurity potential by
\begin{equation}
u_0 = \int U_0(|\mathbf{r}|) \,d\mathbf{r} \:,\;\; u_3 = \int
U_3(|\mathbf{r}|) \,x^2 y^2 z^2 d\mathbf{r} \:;
\end{equation}
$E$ is an energy given by $E=6\bar{\varepsilon}$ if $\hbar\omega
\ll \bar{\varepsilon}$ and $E=\hbar\omega$ if $\hbar\omega \gg
\bar{\varepsilon}$; $Q$ is a parameter, which is determined by the
QW confinement potential and the doping profile,
\begin{equation}\label{Q}
Q = \left. \int\limits_{-\infty}^{+\infty}
\frac{d\:\varphi_1^3(z)\varphi_2(z)}{dz} \, w(z) dz \right/
\int\limits_{-\infty}^{+\infty} \varphi_{1}^4(z) w(z) dz  \:.
\end{equation}
The parameter $Q$ equals to $2\pi/a$ for a rectangular quantum
well with infinitely high barriers where the layer of impurities
is placed in the QW center.

\section{Discussion}

We have shown that the free-carrier absorption of circularly
polarized radiation in quantum well structures leads to an
electric current, which reverses its direction upon switching the
light helicity. The proposed pure orbital mechanism of the
circular photogalvanic effect is based on interference of
different pathways contributing to the light absorption and does
not involve spin of free carriers.

In QW structures grown along [110] or [113] crystallographic
direction the circular photocurrent can be excited at normal
incidence of the radiation. Following Eqs.~(\ref{j_110}) and
(\ref{gamma3}) we can estimate the magnitude of the circular
photocurrent density. It gives $j\sim 2\times10^{-10}\,$A/cm for
the photon energy $\hbar\omega=10\,$meV, the light intensity
$I=1\,$W/cm$^2$, the carrier density $N_e=10^{12}\,$cm$^{-2}$, and
the band parameters
$P/\hbar\approx\sqrt{E_g/(2m^*)}\approx1.3\times10^8\,$cm/s,
$|\Xi_{cv}/\Xi_c|\approx0.36$~\cite{optor}, which are appropriate
to GaAs-based structures.

In (001)-oriented QWs the helicity dependent photocurrent can be
induced at oblique incidence of the radiation only, and the
current magnitude varies in a wide range depending on the QW width
and asymmetry. Equations~(\ref{j_phenom}) and (\ref{gamma1}) give
$j\sim 10^{-8}\,$A/cm for a QW of the width $a=150\,$\AA, the
asymmetry degree $\xi=0.2$, $l_x=0.2$, and the photon energy, the
light intensity and the carrier density presented above. The
estimated magnitude of the photocurrent corresponds to that
measured in experiments on (001)-oriented
structures~\cite{GanichevPrettl03}.

In addition to the orbital mechanism, a contribution to CPGE in
quantum wells structures may come from $\mathbf{k}$-linear
spin-orbit splitting of the quantum subbands together with the
spin-sensitive selection rules for the optical transitions.
However, estimations show that the spin contribution to the
circular photocurrent induced by the free-carrier absorption in
$n$-doped structures vanishes to the first order in spin-orbit
interaction and, therefore, is considerably smaller than the
orbital term.

Additional experiments, such as measurements of the frequency and
temperature dependencies of the photocurrent, can be useful in
establishing the roles of the orbital and spin-related mechanisms
of CPGE more reliable. In particular, the ratio
$\gamma_1/\gamma_2$ is expected to be independent of the photon
energy $\hbar\omega$ for the spin-related mechanism, while it does
depend on $\hbar\omega$ for the orbital mechanism provided
$\hbar\omega> \bar{\varepsilon}$. Besides, increase of the
temperature from liquid helium to the room temperature leads to
changing the dominant scattering mechanism. It can strongly affect
the orbital contribution to CPGE, which is inherently related to
details of scattering.

\paragraph*{Acknowledgments.} The author acknowledges useful
discussions with E.L.~Ivchenko, L.E.~Golub, and N.S.~Averkiev.
This work was supported by the RFBR, programs of the RAS,
President Grant for young scientists, and the Russian Science
Support Foundation.


\begin{thebibliography}{16}
\bibitem{scattering} Impurity potential of the form~(\ref{potential})
can give rise to non-diagonal components of the conductivity
tensor, $\sigma_{xy}$ and $\sigma_{yx}$, in asymmetrical
(001)-grown QWs.
\end{thebibliography}
\end{document}